\documentclass[twocolumn,superscriptaddress, longbibliography,
amsmath,amssymb,
]{revtex4-2}

\usepackage{graphicx}
\usepackage{dcolumn}
\usepackage{bm}
\usepackage{xcolor}

\begin{document}
	
	\preprint{APS/123-QED}
	
	\title{Strong and weak dynamo regimes in Taylor-Couette flows}
	
	\author{Ashish Mishra}
	\email{a.mishra@hzdr.de}
	\affiliation{Helmholtz-Zentrum Dresden-Rossendorf, Bautzner Landstr. 400, D-01328 Dresden, Germany}
	\author{George Mamatsashvili}%
	\affiliation{Helmholtz-Zentrum Dresden-Rossendorf, Bautzner Landstr. 400, D-01328 Dresden, Germany}
	\affiliation{Abastumani Astrophysical Observatory, Abastumani 0301, Georgia}
	\author{Frank Stefani}
	\affiliation{Helmholtz-Zentrum Dresden-Rossendorf, Bautzner Landstr. 400, D-01328 Dresden, Germany}
	
	
	
	
	\date{\today}
	
	\begin{abstract}
		
	We reveal a nonlinear magnetic dynamo in a Taylor-Couette flow at small magnetic  Prandtl numbers $Pm\leq 1$,  which has been previously believed to exist only at higher $Pm\gtrsim 10$ in this flow.  The amplitude of initial perturbations, $Pm$ and domain aspect ratio  play a key role in the onset and evolution of the dynamo.  It exists in two main states -- a weak state dominated by large-scale modes and a strong,  turbulent state with higher amplitude dominated by small-scale modes.  These findings can be important for dynamo processes in various astrophysical objects with small $Pm$.

	\end{abstract}
	
	\maketitle
	
	
	Experimental studies on fundamental magneto-hydrodynamic (MHD) processes in astrophysical disks are usually based on a Taylor-Couette (TC) flow of liquid metals between two coaxial cylinders in the laboratory.  By adjusting the rotation of the cylinders, one can approximately match the rotation profile of the fluid to the Keplerian rotation, $\Omega \propto r^{-3/2}$, of the disks. One of the most important applications of a TC flow subject to an imposed magnetic field is in the studies of different -- standard \cite{Ji_etal2001,Goodman_Ji2002, Ruediger_Zhang2001, Wei_etal2016, Winarto_etal2020, Mishra_etal2022, Wang_etal2022, Mishra_etal2023, Mishra_etal2024}, helical \cite{Hollerbach_Ruediger2005, Stefani_etal2006, Liu_etal2006, Stefani_etal2009, Kirillov_Stefani2010, Mamatsashvili_Stefani2016, Mamatsashvili_etal2018,Mamatsashvili_etal2019} and azimuthal \cite{Hollerbach_etal2010, Seilmayer_etal2014, Mishra_etal2021, Mishra_etal_AMRI_Convec} -- flavors of magnetorotational instability (MRI), which is a central process driving angular momentum and mass transport in the disks (see reviews \cite{Ruediger_etal2018, Ji_Goodman2023, Stefani2024}). 
	
	MRI is closely interlaced with magnetic dynamo, leading to the concept of the nonlinear MRI-dynamo -- an important class of instability-driven processes of amplification and sustenance of a large-scale magnetic field, which, in turn, supports that instability \cite{Rincon2019}. The MRI-dynamo has been extensively studied mainly in the local shearing-box model of the disks,  which allowed to clarify its onset, sustenance dynamics and dependence on fluid viscosity $\nu$ and magnetic diffusivity $\eta$ \cite{Rincon_etal2007, Lesur_Ogilvie2008, Herault_etal2011, Riols_etal2015, Walker_etal2016, Walker_etal2017, Shi_etal2016, Nauman_Pessah2016, Riols_etal2017, Mamatsashvili_etal2020, Held_Mamatsashvili2022, Guilet_etal2022}. In particular, this dynamo was shown to be most sensitive to the magnetic Prandtl number $Pm=\nu/\eta$, being sustained at $Pm\gtrsim Pm_c$ and non-existent at $Pm\lesssim Pm_c$, where the threshold  $Pm_c \sim 1$ depends on the system parameters, such as box aspect ratio, flow shear, etc.  
	
	In parallel with local,  there have been global studies of dynamo in TC flows to understand magnetic field generation in disks beyond the local analysis.  Early works considered TC flows hydrodynamically unstable due to Rayleigh's centrifugal criterion and showed that dynamo in those flows is kinematic,  being governed by the specific flow structure (Taylor vortices) resulting from the hydrodynamic instability \cite{Willis_Barenghi2002,Laguerre_etal2008, Nore_etal2012,Gissinger2014, Marcotte_etal2021}.  
	
	Distinct from this, Guseva et al. \cite{Guseva_etal2017} considered quasi-Keplerian TC flows and demonstrated a sustained subcritical dynamo action only at high enough $Pm \gtrsim 10$ and Reynolds numbers $Re\sim 10^4$.  Since a quasi-Keplerian TC flow itself is Rayleigh-stable \cite{Ji_etal2006, Avila2012, Shi_etal2017} and hence unable to give rise to a kinematic dynamo,  this dynamo represents instead a complex type of nonlinear, or finite-amplitude dynamo, which,  in contrast to the kinematic one,  is excited at sufficiently large initial perturbations of velocity or magnetic field and relies on MRI for energy supply.  As a result,  the flow settles down into MHD turbulence with an associated large-scale field that mutually sustain each other via the interplay of MRI and nonlinear feedback, as in the local case. This dynamo is in fact neither purely large- nor small-scale but involves a range of wavelengths.  In this regard,  another type of a dynamo process in a quasi-Keplerian TC flow driven by Tayler instability was identified at sufficiently high $Rm$ (and hence high $Pm$) in \cite{Ruediger_Schultz2020}, which, however, can exist even at $Pm<1$ in spherical geometry \cite{Petitdemange_etal2023}.
	
	In this Letter, we report a nonlinear dynamo action persisting at small $Pm \leq1$ in a Rayleigh-stable quasi-Keplerian TC flow and characterize its properties. Depending on the amplitude of initial perturbations and $Pm$, this dynamo can exist in the weak and strong states classified according to the magnetic energy of axisymmetric versus non-axisymmetric modes. 
	

	We consider a standard TC setup consisting of inner and outer cylinders with radii $r_{in}$ and $r_{out}=2r_{in}$ and height $L_z$ rotating with angular velocities $\Omega_{in}$ and $\Omega_{out}$, respectively, in the cylindrical coordinate system $(r, \phi, z)$.  The cylinders are tall with a vertical aspect ratio $L_z/r_{in}=10$.  The rotation ratio is  $\Omega_{out}/\Omega_{in}=0.35$ corresponding to the Rayleigh-stable quasi-Keplerian rotation of the fluid between the cylinders \cite{Ruediger_etal2018}, so that dynamical processes are only of MHD nature.  The non-ideal MHD equations for an incompressible fluid are
	\begin{equation} \label{equation_momentum}
		\frac{\partial {\boldsymbol u}}{\partial t}+({\boldsymbol  u}\cdot\nabla){\boldsymbol u}=-\frac{1}{\rho} \nabla p + \frac{(\nabla\times\boldsymbol{B})\times {\boldsymbol B}}{\mu_0\rho}+\nu \nabla^2 {\boldsymbol u},
	\end{equation}
	\begin{equation}	\label{equation_induction}
		\frac{\partial \boldsymbol{B}}{\partial t}=\nabla \times(\boldsymbol{u}\times \boldsymbol{B})+\eta \nabla^2 \boldsymbol{B}	
	\end{equation}
\noindent
	together with the divergence-free conditions $\nabla \cdot \boldsymbol{u}=\nabla \cdot \boldsymbol{B}=0$,  where $\rho$ is the constant density, $\boldsymbol{u}$ is the velocity, $p$ is the pressure and $\boldsymbol{B}$ is the magnetic field.  $\mu_0$ is the vacuum magnetic permeability.  We normalize length by $r_{in}$, time by $\Omega_{in}^{-1}$, ${\boldsymbol u}$ by $\Omega_{in} r_{in}$, $p$ by $\rho r_{in}^2\Omega_{in}^2$ and ${\boldsymbol B}$ by $\Omega_{in} r_{in}\sqrt{\rho\mu_0}$. The main parameters are the Reynolds, $Re=\Omega_{in} r_{in}^2/\nu$, magnetic Reynolds, $Rm=\Omega_{in} r_{in}^2/\eta$, and magnetic Prandtl, $Pm=Rm/Re$, numbers. We fix $Re=10^5$ to be an order higher than that used  in a related work \cite{Guseva_etal2017}, but vary $Pm$.

	\begin{figure}
		\includegraphics[width=0.9\columnwidth]{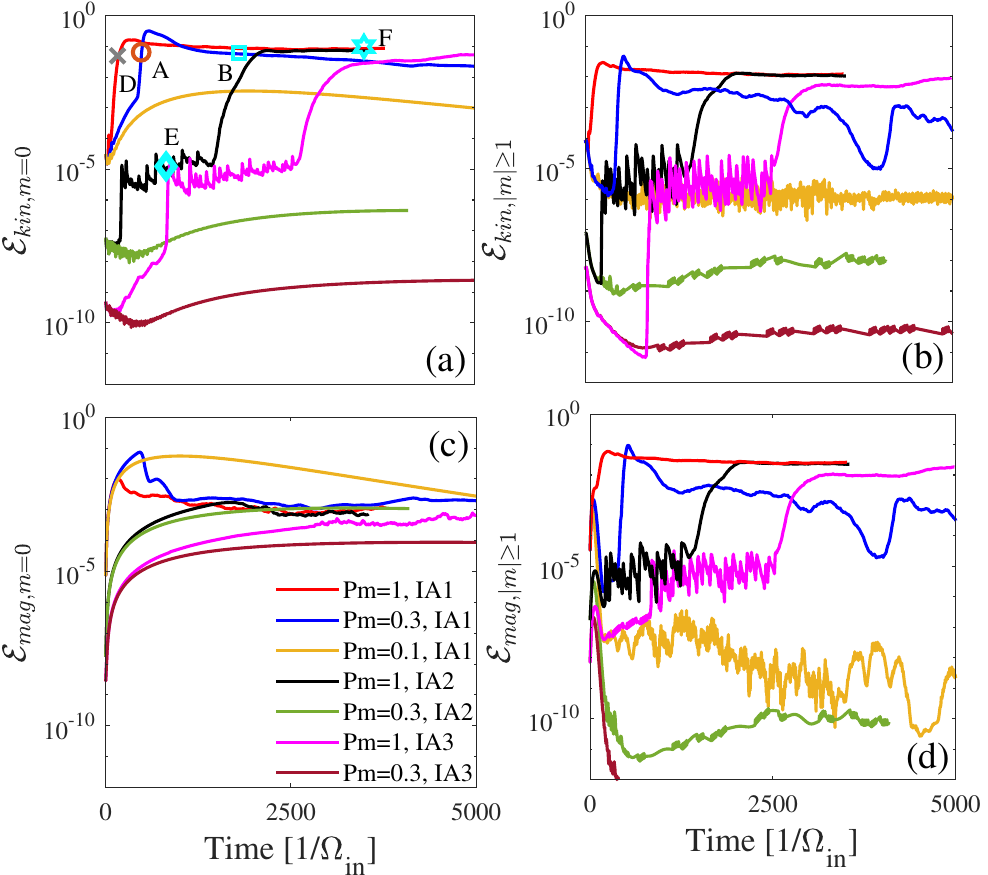}
		\caption{Evolution of the volume-integrated kinetic $\mathcal{E}_{kin}$ (top) and magnetic $\mathcal{E}_{mag}$ (bottom) energies for (a,c) axisymmetric $m=0$ and (b,d) non-axisymmetric $|m|\geq 1$ modes for different initial amplitudes (IAs) and $Pm \in \{0.1, 0.3, 1\}$.} \label{Amp_dependence}
	\end{figure}

	To solve Eqs. (\ref{equation_momentum}) and (\ref{equation_induction}), we use the pseudo-spectral code described in \cite{Guseva_etal2017}.  It employs a high-order finite-difference method along radius $r$ and Fourier expansion in the azimuthal $\phi$- and axial $z$-directions. We take $N_r=640$ radial points and $N_z=640$ Fourier modes in the axial direction.  The total number of azimuthal modes $|m|\leq N_{\phi}$, where $m$ is the azimuthal wavenumber, is set to $N_{\phi}=32$.  These resolutions are sufficient to reach convergence of the results for the adopted parameters, as shown by the resolution test in Fig. \ref{spectra_resolution_study} of End Matter.  Boundary conditions are no-slip for the velocity and insulating for the magnetic field at the cylinder walls and periodic along the $z$-axis with a spatial period of $L_z$. Initial conditions are random perturbations of velocity and magnetic field with the same rms initial amplitude (IA) on top of the TC flow without a mean external field.  Since the dynamo is nonlinear (subcritical), we take different $\rm IA3 : IA2 : IA1 = 1:4:65$ to explore its dependence on the magnitude of initial perturbations. Simulations have been run up to $10^4\Omega_{in}^{-1}$, about 7 times longer than in \cite{Guseva_etal2017} to better see the long-time behavior (sustenance) of the dynamo.
	
	\begin{figure}
		\includegraphics[width=0.9\columnwidth]{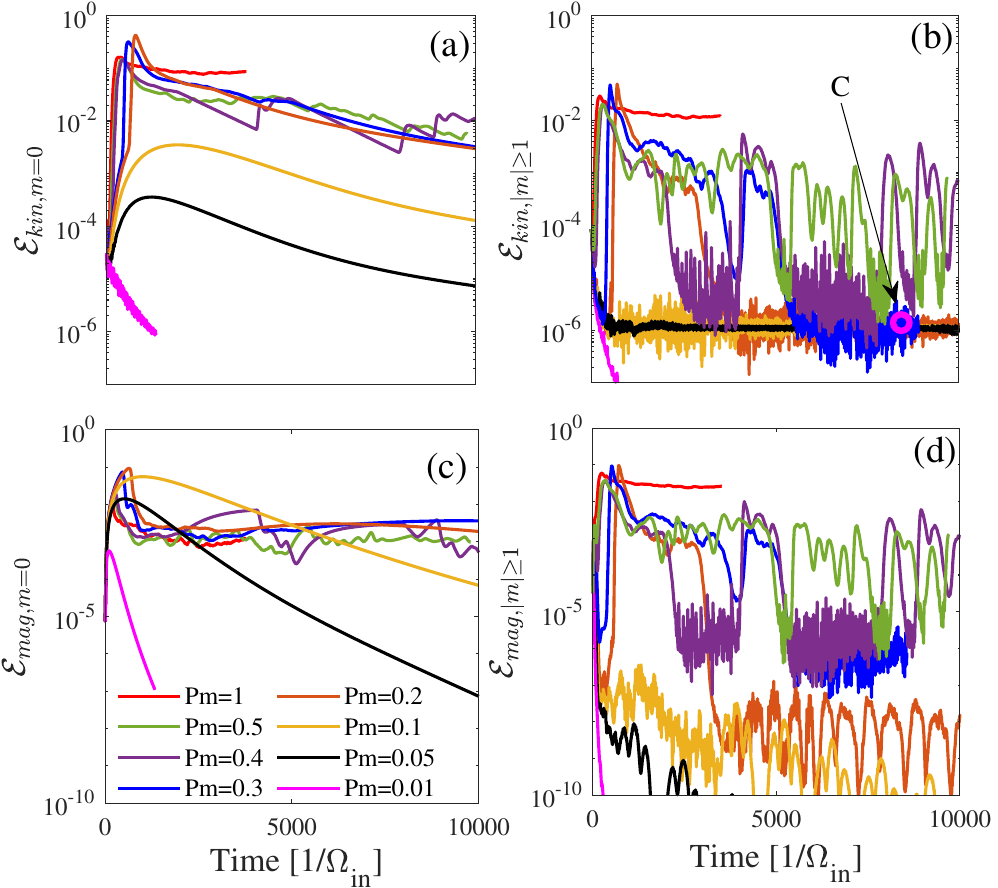}
		\caption{Same as in Fig. \ref{Amp_dependence} but only for the largest initial amplitude IA1 and a broader range of $Pm \in \{0.01, 0.05, 0.1, 0.2, 0.3, 0.4, 0.5, 1\}$.  The structure of the weak dynamo state at time moment C in (b) is depicted in Fig. \ref{rzslices}.} \label{Pm_dependence}
	\end{figure}
		\begin{figure*}
		\includegraphics[width=0.9\textwidth]{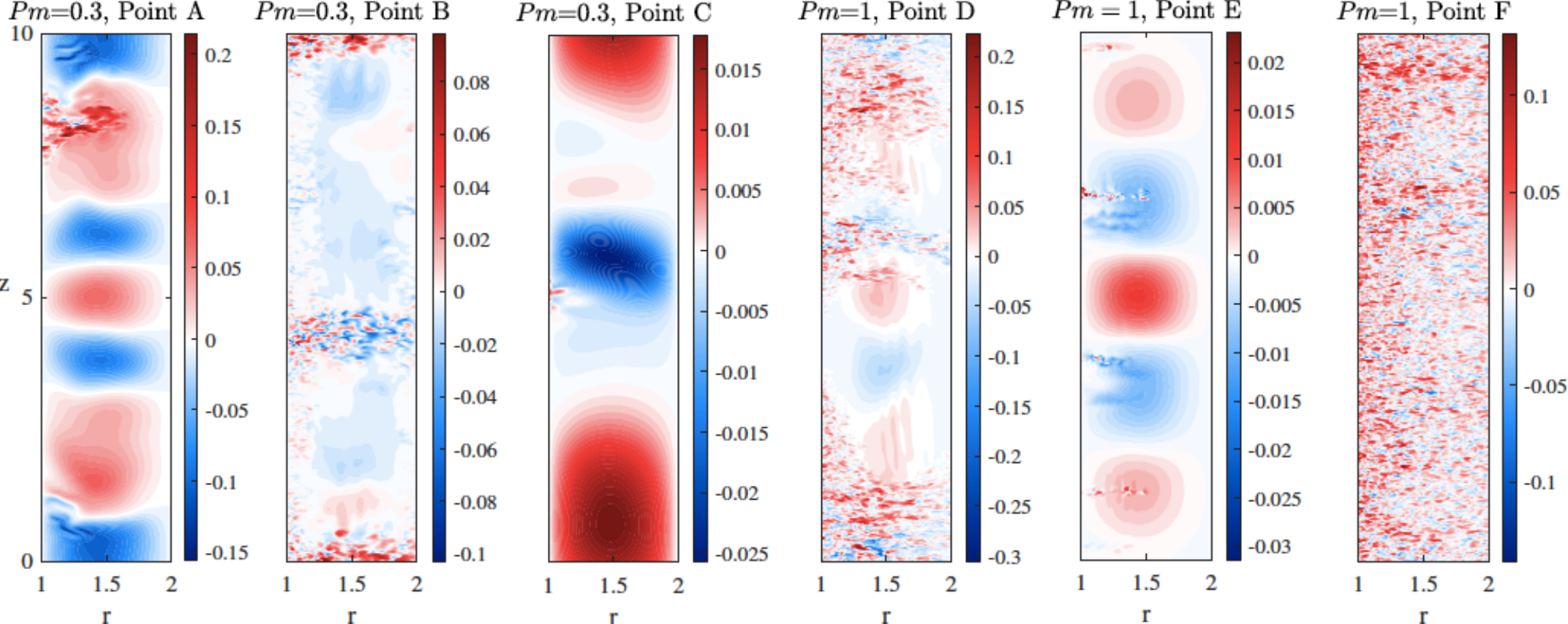}
		\caption{Structure of the azimuthal field $B_{\phi}$ in the $(r,z)$-plane at a given azimuthal angle $\phi$ and the characteristic time moments A, B, C for $Pm=0.3$ and D, E, F for $Pm=1$ from Figs. \ref{Amp_dependence} and \ref{Pm_dependence}, which represent different phases and regimes of the dynamo at these $Pm$.  Specifically,  the moments A and D correspond to the growth stages, while the moments B and F to the strong and moments C and E to the weak dynamo states.} \label{rzslices}
	\end{figure*}

	 Figure \ref{Amp_dependence} shows the evolution of the volume-integrated non-dimensional kinetic, $\mathcal{E}_{kin}=\boldsymbol{u}^2/2$ (with the base TC flow subtracted), and magnetic, $\mathcal{E}_{mag}=\boldsymbol{B}^2/2$, energy densities of axisymmetric $m=0$ and non-axisymmetric $|m|\geq 1$ modes for different $Pm \in \{0.1, 0.3, 1\}$ and the initial perturbation amplitudes IA1, IA2 and IA3. Time moments A, B, D, E and F in Fig. \ref{Amp_dependence}(a) denote the key stages of the dynamo evolution depending on IA and $Pm$. For $Pm=1$ and the largest IA1, the magnetic and kinetic energies first increase exponentially in time (moment D) due to MRI and directly reach a saturated quasi-steady state (moment F), while at smaller IA2 and IA3, these energies first reach an intermediate state with about the same low energy level (moment E), which lasts for a while, and afterwards they grow again, saturating at the high level similar to that for IA1.  In this high energy state where the dynamo eventually arrives, the kinetic energy of axisymmetric modes is about an order of magnitude higher than that of non-axisymmetric ones, whereas the magnetic energy of axisymmetric modes is about an order of magnitude smaller than that of non-axisymmetric ones. Both kinetic and magnetic energies of all the modes are nearly constant in time. We refer to this state, where magnetic energy is high and mostly contained in the non-axisymmetric modes, as the \emph{strong} dynamo state. On the other hand, in the intermediate low energy state, kinetic energies of axisymmetric and non-axisymmetric modes rapidly fluctuate and are comparable, whereas the magnetic energy of axisymmetric modes varies smoothly and is an order of magnitude larger than that of non-axisymmetric ones, which rapidly fluctuates. This state, where magnetic energy is lower and mostly contained in the axisymmetric modes, is referred to as the \emph{weak} dynamo state.

    At lower $Pm=0.3$, the strong dynamo state occurs only at the largest IA1 (moment B),  which sets in just after the MRI growth phase (moment A).  However, after some time the dynamo settles down in the weak state, as seen in Fig. \ref{Pm_dependence}. For smaller IA2, the dynamo is always in the weak state with the magnetic energy of axisymmetric modes being orders of magnitude higher than their kinetic energy, while for non-axisymmetric modes kinetic energy is much larger than their magnetic energy. For IA3 the magnetic energy decays and hence there is no dynamo. At even lower $Pm=0.1$, there is some initial growth of the magnetic energy that decays later (Fig. \ref{Pm_dependence}), so also this case is not a dynamo.  Note that the saturation level of the kinetic energy of the non-axisymmetric modes for IA3 and IA2 at $Pm=1$ matches that for IA1 at $Pm=0.1$,  although the magnetic energy decreases later.  This similarity may suggest the existence of the lowest energy, marginal state separating non-dynamo and dynamo regimes.

Figure \ref{Pm_dependence} shows the evolution of the energies for the largest IA1 and a broader range of $Pm \in \{0.01, 0.05, 0.1, 0.2, 0.3, 0.4, 0.5, 1\}$.  In this case, both dynamo states occur for $Pm \gtrsim 0.1$, while for $Pm \lesssim 0.1$ the magnetic energy decays on the ohmic timescale and hence there is no dynamo.  For $Pm=\{0.2, 0.3\}$, the strong dynamo initially lasts for a while but eventually falls into the weak state, for $Pm=\{0.4, 0.5\}$ the dynamo irregularly vacillates between the strong and weak states, as seen in the evolution of the non-axisymmetric mode energies in Figs.  \ref{Pm_dependence}(b) and \ref{Pm_dependence}(d). By contrast, a statistically steady strong dynamo is consistently observed at $Pm=1$.

Figure \ref{rzslices} shows the structure of the azimuthal field $B_{\phi}$ in the $(r,z)$-plane in both dynamo states at the characteristic moments A, B, C for $Pm=0.3$ and D, E, F for $Pm=1$ marked in Figs. \ref{Amp_dependence} and \ref{Pm_dependence}. During a rapid growth phase for $Pm=0.3$ (moment A), the azimuthal field has the form of regular large-scale cells, like those of MRI under a constant axial background field \cite{Winarto_etal2020, Mishra_etal2022, Wang_etal2022} with the small turbulent spots emerging at the inner cylinder, while for $Pm=1$ (moment D) a larger portion of the flow is turbulent, coexisting with the regular cell structures. In the later strong state for $Pm=1$ (moment F), this turbulence dominated by small-scale structures finally fills up the entire domain, whereas for $Pm=0.3$ (moment B) the flow consists of evenly distributed turbulent spots between more regular larger-scale structures.  These large-scale cells do not fully fade away,  indicating continual competition between them and localized turbulence. On the other hand, in the weak states both for $Pm=0.3$ (moment C) and $Pm=1$ (moment E), these cells are the dominant structures with some narrower spots of weaker turbulence. Thus,  comparing the field structures  in Fig. \ref{rzslices}, we notice that the transition from the weak to strong dynamos with increasing $Pm$ occurs via the emergence and growth of turbulent spots on the regular cells of the weak dynamo.  These turbulent spots spread and prevail over the cells, filling up the flow domain in the strong dynamo state.  

\begin{figure}
        \vspace{-2em}
        \centering
		\includegraphics[width=0.65\textwidth]{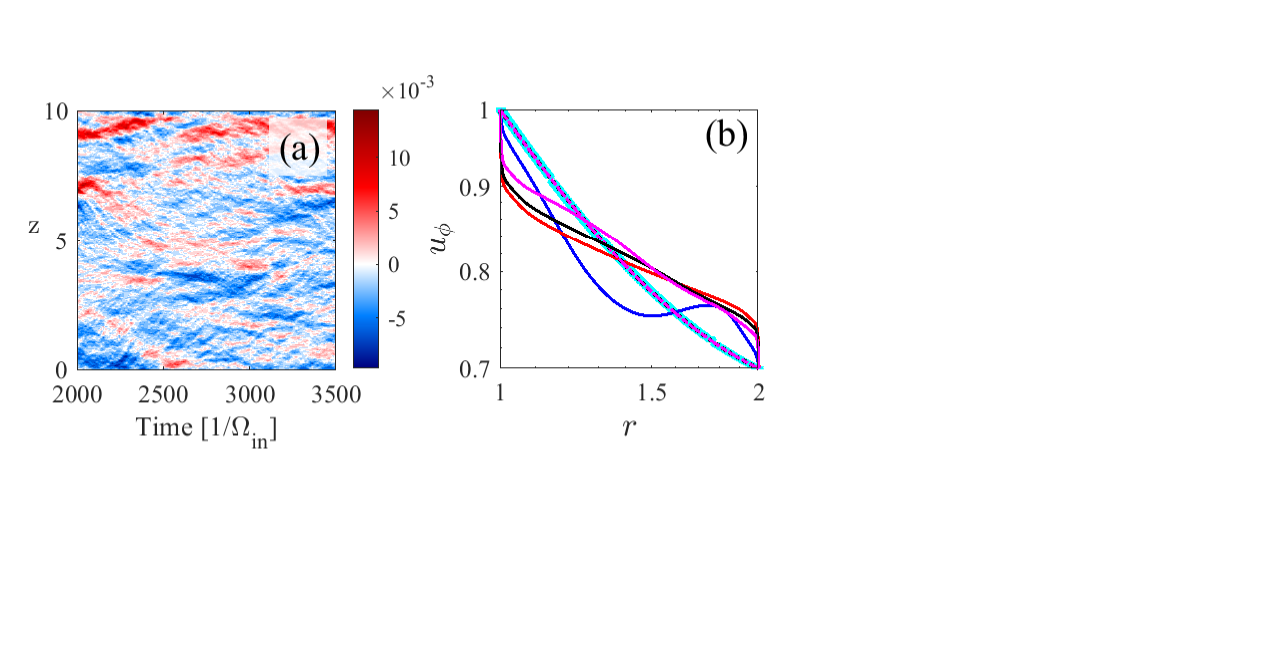}
		\vspace{-7em}
		\caption{(a) Space-time diagram of the radially- and azimuthally-averaged azimuthal field $\langle B_\phi\rangle_{r,\phi}$ for the run with $Pm = 1$ and IA1. (b) Time-averaged radial profiles of $u_{\phi}$ established in the  strong and weak dynamo states at mid-height $z=5$ for the same cases and color coding as in Fig. 1, where the cyan curve corresponds to the classical TC profile. } \label{Bphi_av_uphi}
	\end{figure}

\begin{figure}
    \centering
    \includegraphics[width=0.45\textwidth]{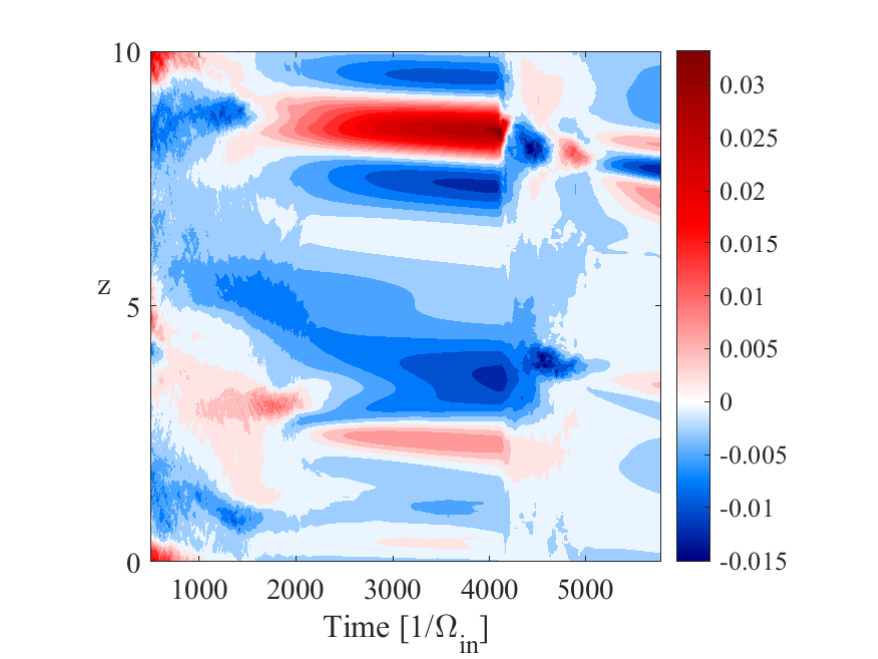}
    \includegraphics[width=0.45\textwidth]{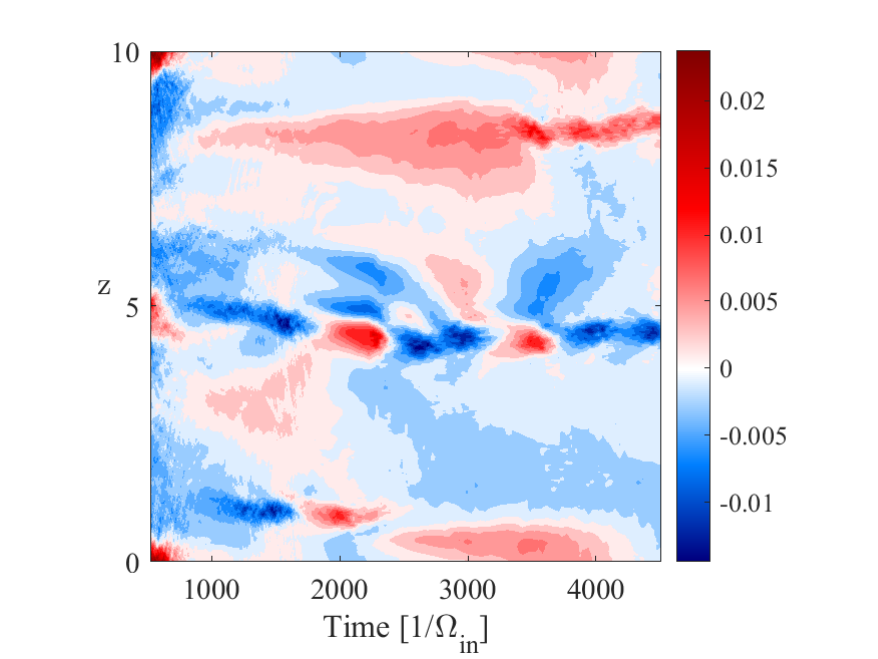}
\caption{Space-time diagrams of the radially- and azimuthally-averaged azimuthal field $\langle B_\phi\rangle_{r,\phi}$ represented as a function of time and $z$ for $Pm=0.4$ (left), $Pm=0.5$ (right) and the largest IA1. This  mean azimuthal field has larger axial length- and time-scales in the weak dynamo state, which both decrease in the strong state, as seen for the $Pm=1$ case in Fig. \ref{Bphi_av_uphi}(a).  Nevertheless, the mean field, changing over much longer times  than the dynamical/orbital time $\sim \Omega_{in}^{-1}$ -- a characteristic time of MRI, serves as a supporting background field for MRI to grow.} \label{contour_bphi_time}
\end{figure}

Regarding a mechanism driving the dynamo, in the related local shearing box models of a Keplerian flow characterized, as the present setup, by  zero net magnetic flux, turbulent MRI-dynamo is sustained due to the interplay between the nonmodal, or transient growth of MRI for non-axisymmetric modes and nonlinearity (transverse cascade), with the latter providing a positive feedback to the former  \cite{Lesur_Ogilvie2008, Herault_etal2011, Riols_etal2017, Gogichaishvili_etal2017, Mamatsashvili_etal2020, Held_Mamatsashvili2022}. These two main processes should also be at work in the considered global TC flow. Specifically, energy is extracted from the TC flow into perturbations by MRI, which is supported by the slowly-varying (compared to the dynamical time $\sim\Omega_{in}^{-1}$), azimuthally- and radially-averaged component of the dominant azimuthal field, $\langle B_{\phi}\rangle_{\phi,r}$, whose dependence on time and $z$ is shown in Fig. \ref{Bphi_av_uphi}(a) (see also Fig.  \ref{contour_bphi_time} in End Matter). This MRI-supporting mean azimuthal field is in turn replenished by the nonlinear interaction of the MRI-amplified non-axisymmetric modes. So, this interplay between MRI and the large-scale field reinforcing each other is the essence of the found nonlinear dynamo.  The nonlinear saturation of MRI in TC flows occurs via the deviation of the radial profile of the mean azimuthal velocity, $u_{\phi}$,  from the classical TC profile,  so as to reduce the shear of the bulk flow, thereby halting MRI growth \cite{Liu_etal2006, Mishra_etal2023}.  Here this deviation of $u_{\phi}$ is more pronounced in the strong dynamo state [Fig. \ref{Bphi_av_uphi}(b)].

\begin{figure}
		\includegraphics[width=\columnwidth]{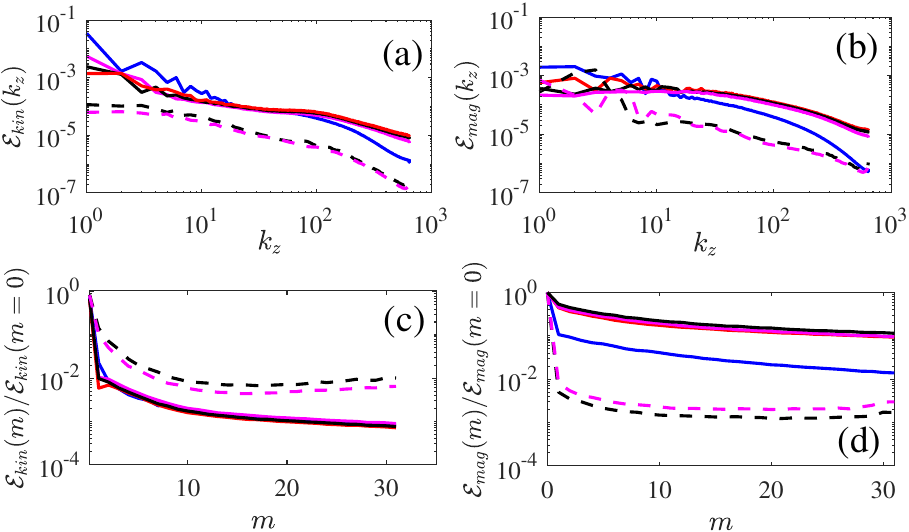}
		\caption{Kinetic (left) and magnetic (right) energy spectra as a function of (a,b) the axial wavenumber $k_z$ and (c,d) azimuthal wavenumber $m$ in the strong (solid curves) and weak (dashed curves) dynamo states for the same parameters (colors of the curves) as in Fig. \ref{Amp_dependence}. The azimuthal spectra are normalized by the corresponding energies of the axisymmetric $m=0$ mode.} \label{spectra}
	\end{figure}

\begin{figure}[h]
    \centering
    \includegraphics[width=0.45\textwidth]{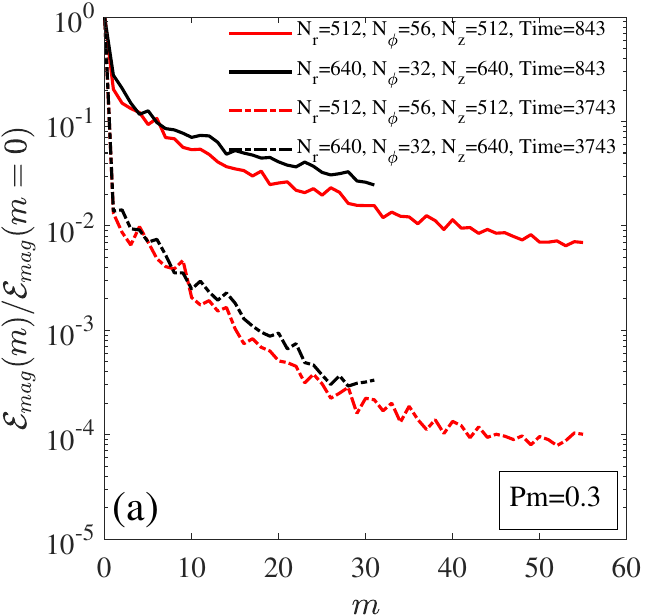}
    \hspace{2em}
    \includegraphics[width=0.45\textwidth]{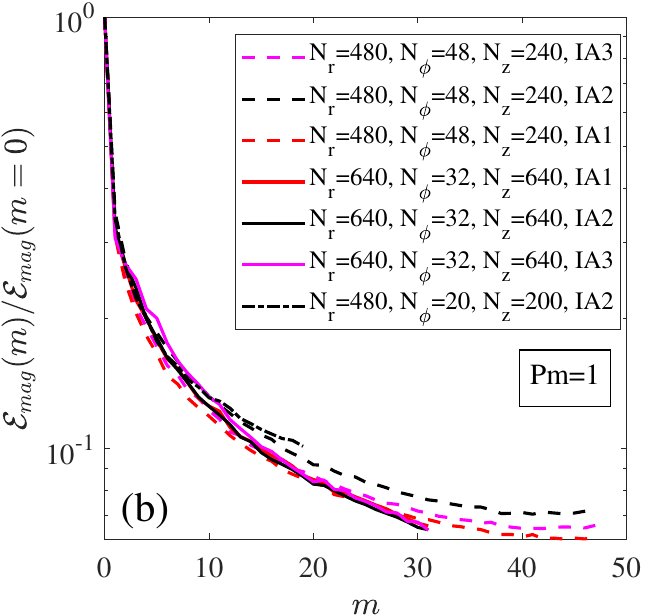}
\caption{Magnetic energy spectra as a function of $m$ normalized by the energy of the axisymmetric $m=0$ mode [see Fig. \ref{spectra}(d)] obtained for different radial $N_r=480, 512, 640$, azimuthal $N_{\phi}=20, 32, 48, 56$ and axial $N_z=200, 240, 512, 640$ resolutions (a) at different times for $Pm=0.3$ and the largest IA1 and (b) averaged in time just after the saturation for $Pm=1$ and different IA. This resolution study demonstrates that the spectra well converge, validating the resolution $(N_r,N_{\phi},N_z)=(640,32,640)$ used in the main analysis.} \label{spectra_resolution_study}
\end{figure}

\begin{figure}
    \centering
    \includegraphics[width=0.7\textwidth]{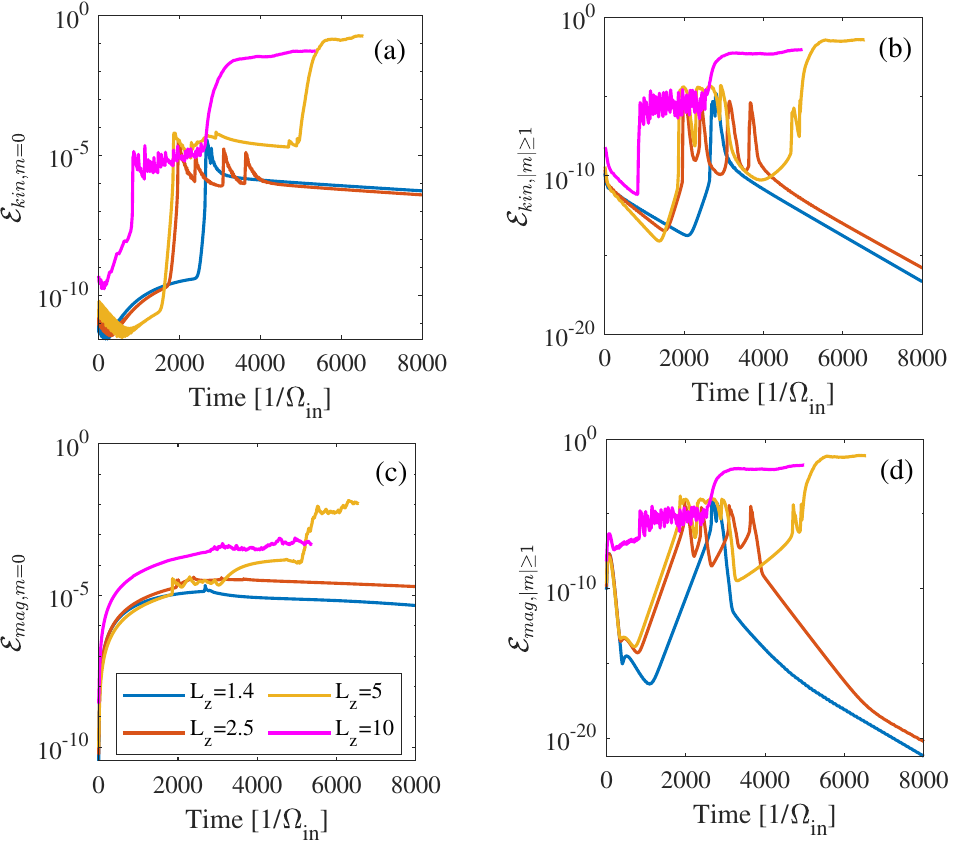}
\caption{Evolution of the volume-integrated kinetic, $\mathcal{E}_{kin}$, and magnetic, $\mathcal{E}_{mag}$, energies for (a,c) axisymmetric and (b,d) only non-axisymmetric modes for different height $L_z\in \{1.4, \, 2.5, \, 5,\, 10\}$ of the domain at $Pm =1$ and IA3. Simulations for $L_z \in \{1.4, \, 2.5, \, 5\}$ were conducted with resolution $(N_r, \, N_{\phi}, \, N_z ) =(480, 16, 240)$ while for $L_z=10$ with $(N_r, \, N_{\phi}, \, N_z )=(640, 32, 640)$. It is seen that for $L_z\in \{1.4, \, 2.5\}$ the energies grow transiently, then decay more rapidly for non-axisymmetric modes and much slower (i.e., at times larger than the final time $8000$ in these plots) for axisymmetric ones due to dissipation, failing to exhibit a sustained dynamo. By contrast, the cases with larger $L_z\in \{5, \, 10\}$ are clearly a dynamo -- the energies of both axisymmetric and non-axisymmetric modes rapidly grow several orders of magnitude and saturate in the strong state.}      \label{diff_Lz}
\end{figure}

	To find out dynamically important length-scales,  we computed kinetic and magnetic energy spectra  as a function of the axial $k_z$ and azimuthal $m$ wavenumbers, integrating them along radius $r$.  Then, these spectra were averaged in time separately in the strong and weak dynamo states. The time-averaged kinetic and magnetic energy spectra in both states summed over $m$ are plotted as a function of $k_z$, respectively, in Figs. \ref{spectra}(a) and \ref{spectra}(b), while the same spectra integrated along $k_z$ are plotted as a function of $m$, respectively, in Figs. \ref{spectra}(c) and \ref{spectra}(d). It is seen that the main qualitative distinction between these two states is that in the strong state the magnetic energy spectrum spreads over a broad range of $k_z$, from the lowest one $2\pi/L_z$ set by the domain height $L_z$  up  to the highest resistive one $Rm^{1/2}$,  whereas in the weak state, it is concentrated mostly at low to intermediate $k_z \lesssim 10$ and is steeper [Fig. \ref{spectra}(b)]. In the strong state, the magnetic spectrum  is nearly flat up to $k_z \sim 20$ for $Pm=1$ and for all IA, but drops at larger $k_z$ due to dissipation.  Also, it becomes steeper with decreasing $Pm$,  as seen from the  $Pm=0.3$ case in this figure. The kinetic energy spectrum  behaves similarly in the strong state, but it is higher at lower $k_z\lesssim 10$, having at these $k_z$ more power at $Pm=0.3$ than that at $Pm=1$ [Fig. \ref{spectra}(a)]. At intermediate  $10 \lesssim k_z \lesssim 100$, it is flatter for $Pm=1$ than that for $Pm=0.3$ and then decreases in the dissipation range at higher $k_z\gtrsim 100$ the steeper the smaller $Pm$ is. On the other hand, in the weak dynamo state at $Pm=1$, the kinetic energy spectrum is mostly flat at $k_z\lesssim 10$ but drops more steeply with increasing $k_z$ than that in the strong state.  Thus,  the larger domain height ($L_z/r_{in}=10$) adopted here has in fact allowed us to uncover the large-scale properties of the dynamo at lower $k_z$, such as the increased kinetic energy, i.e., large-scale motions, and weak state compared to those described for shorter domains with $L_z/r_{in}=1.4$ in \cite{Guseva_etal2017}.  More importantly, the taller domain and an order higher $Re=10^5$ in our case are those key differences from Ref.  \cite{Guseva_etal2017} that enabled the existence of the dynamo at lower $Pm\leq 1$. This is further confirmed by analysis of the $L_z$ dependence (Fig. \ref{diff_Lz} in End Matter) and is consistent with local studies of MRI-dynamo \cite{Nauman_Pessah2016, Shi_etal2016, Walker_etal2017, Held_Mamatsashvili2022, Guilet_etal2022}. 
	
	The azimuthal spectra in Figs. \ref{spectra}(c) and \ref{spectra}(d) are normalized by the corresponding energies of the $m=0$ mode to highlight the role of the non-axisymmetric $|m|\geq 1$ modes relative to the axisymmetric one.  In the strong state, the kinetic energy for $m=0$ is orders of magnitude higher than that for $|m|\geq 1$, but the contribution of the latter modes increases in the weak state [Fig. \ref{spectra}(c)].  By contrast, the magnetic energy for $|m|\geq 1$ is comparable to, but still smaller than that for $m=0$, and it further decreases relative to that of the $m=0$ mode in the weak state [Fig. \ref{spectra}(d)], implying the field is nearly axisymmetric. The resolution study and convergence of these magnetic $m$-spectra  are shown in Fig. \ref{spectra_resolution_study} of End Matter.


In this Letter, we revealed a subcritical dynamo in a quasi-Keplerian TC flow at low magnetic Prandtl $Pm\leq 1$ and high Reynolds $Re=10^5$ numbers. This basic flow resembling astrophysical disks has been previously believed \emph{not} to support dynamo at small $Pm$, since it is generally hard to trigger dynamo at such $Pm$ in shear  flows \cite{Riols_etal2015,Guseva_etal2017, Rincon2019, Nauman_Pessah2016, Mamatsashvili_etal2020}.  Nevertheless, our results indicate that a nonlinear dynamo is possible in the small-$Pm$ regime if $Re$ and initial perturbation amplitude are high enough and the flow domain is sufficiently tall.  The dynamo is driven by MRI and depends on  $Pm$. It exists at $Pm\gtrsim 0.1$ in two main -- weak and strong -- regimes with low and high energy levels, respectively, while there is no dynamo at $Pm\lesssim 0.1$. These strong and weak dynamo states occur for a sufficiently large initial amplitude, respectively, at $Pm \sim 1$ and $Pm\sim 0.2$, while for $0.2 \lesssim Pm \lesssim 1$ the dynamo irregularly vacillates between these two states.  The strong state is fully turbulent dominated by small-scale structures, whereas the weak state is dominated by large-scale structures coexisting with localized turbulence.

These findings can be important for (small-scale) dynamo processes in magnetized  astrophysical flows with strong differential rotation (shear), very high $Re\gtrsim 10^{11}$ and small $Pm\lesssim 1$. Typical examples of such flows are dense and cold interiors (dead zones) of Keplerian protoplanetary disks \cite{Lesur_etal2023},  but can also include mid-latitude regions of the solar tachocline \cite{Tobias_etal2007} or radiative zones in stars \cite{Jouve_etal2020, Meduri_etal2024}.  Although boundary conditions in disks and stars are quite different from those in the wall-bounded TC flows used here, we expect qualitatively similar shear-induced dynamics of the dynamo in these objects. The strong and weak dynamo states found here and irregular alternations between them can have implications for mass transport variabilities in the above-mentioned objects (e.g., episodic accretion onto protostars). Further studies will analyze the sustenance mechanism of the dynamo at small $Pm$ as well as the effects of varying curvature $r_{in}/r_{out}$, rotation ratio $\Omega_{out}/\Omega_{in}$ and $Re$, especially at higher $Re\gtrsim 10^5$.

This work is supported by the Deutsche Forschungsgemeinschaft (DFG) with Grant No. MA10950/1-1 
and Shota Rustaveli National Science Foundation of Georgia (SRNSFG) [grant No. FR-23-1277]. We thank A. Guseva for the code.
	
	
	\bibliography{references}
	
\clearpage
\onecolumngrid
\begin{center}
    {\bf End Matter}
\end{center}
\vspace{1em}





\end{document}